\documentclass[journal]{IEEEtran}

\usepackage{cite}
\usepackage{amsmath}
\usepackage{graphicx}
\usepackage{dblfloatfix}
\usepackage[numbers,sort&compress]{natbib}

\usepackage{flushend}
\usepackage{soul}
\usepackage{color}
\sethlcolor{yellow}

\newcommand{\cell}[4]{\setlength{\tabcolsep}{2pt}\begin{tabular}{lr}TP:&#1\%\\TN:&#2\%\\FP:&#3\%\\FN:&#4\%\end{tabular}}

\begin{document}

\title{Deep Learning for Plasma Tomography and Disruption Prediction from Bolometer Data}

\author{Diogo R. Ferreira, Pedro J. Carvalho, Hor\'{a}cio Fernandes, and JET Contributors%
\thanks{D. R. Ferreira and H. Fernandes are with the Institute for Plasmas and Nuclear Fusion (IPFN), a member of the EUROfusion Consortium, at IST, University of Lisbon, Portugal.}%
\thanks{P. J. Carvalho  is currently with JET at Culham Science Centre, Abingdon, OX14 3DB, UK.}
\thanks{For JET Contributors, see the author list of ``Overview of the JET preparation for deuterium--tritium operation with the ITER like-wall'' by E. Joffrin et al., in \emph{Nuclear Fusion}, vol. 59, no. 11, p. 112021, August 2019.}}

\markboth{IEEE Transactions on Plasma Science \MakeLowercase{(preprint)}}{Ferreira \MakeLowercase{\textit{et al.}}: Deep Learning for Plasma Tomography and Disruption Prediction from Bolometer Data}

\maketitle

\begin{abstract}
The use of deep learning is facilitating a wide range of data processing tasks in many areas. The analysis of fusion data is no exception, since there is a need to process large amounts of data collected from the diagnostic systems attached to a fusion device. Fusion data involves images and time series, and are a natural candidate for the use of convolutional and recurrent neural networks. In this work, we describe how CNNs can be used to reconstruct the plasma radiation profile, and we discuss the potential of using RNNs for disruption prediction based on the same input data. Both approaches have been applied at JET using data from a multi-channel diagnostic system. Similar approaches can be applied to other fusion devices and diagnostics.
\end{abstract}


\section{Introduction}

Deep learning~\cite{lecun15deep} has become the state-of-the-art approach to many problems, especially those related to image processing and natural language processing. Convolutional neural networks (CNNs) have been extremely successful in image classification~\cite{krizhevsky12imagenet,he15delving}, image segmentation~\cite{noh15learning,ronneberger15unet} and object detection~\cite{ren15faster,he16residual}, to cite only a few examples. On the other hand, recurrent neural networks (RNNs) have been used for speech recognition~\cite{graves05framewise,graves13speech}, language modeling~\cite{mikolov10recurrent,sundermeyer12lstm} and machine translation~\cite{sutskever14sequence,bahdanau15neural}, among other applications.

In general, it could be said that CNNs are appropriate for problems involving images and computer vision, whereas RNNs are especially useful for text and other sequential data, including time series~\cite{pankaj15long}. However, this distinction is not clear-cut since, for example, it is possible to analyze images with RNNs~\cite{oord16pixel}, it is possible to perform sequence learning with CNNs~\cite{gehring17convolutional}, and there are hybrid models combining features from both CNNs and RNNs~\cite{shi15convlstm}.

For the purpose of this work, we will focus on two broad categories: (1) image processing for plasma tomography, which will be addressed with CNNs, and (2) time series analysis for disruption prediction, which will be addressed with RNNs.
In both cases, it becomes apparent that deep learning can play a key role in the compute-intensive tasks associated with the processing of large amounts of data originating from fusion diagnostics.

Here, we focus on JET (Joint European Torus), a D-shaped tokamak with a major radius of 2.96 m and a minor radius of 1.25--2.10 m. JET has a vast assortment of diagnostics, including magnetic coils to measure plasma current and instabilities, interferometers and reflectometers to measure plasma density, Thomson scattering to determine the electron temperature, spectroscopy to measure ion temperature, and X-ray cameras to measure electromagnetic radiation, among others.

In this work, we will be using on a specific diagnostic, the bolometer system~\cite{huber07upgraded}, which measures the plasma radiation on a poloidal cross-section of the fusion device. The signals collected from the bolometer system can be used to monitor the plasma state across an entire pulse. Several phenomena, such as impurity transport and accumulation at the plasma core, can be detected from the bolometer signals. Since these impurity-related phenomena are one of the most frequent precursors of disruptions at JET, this diagnostic plays an important role in disruption studies as well.

The data coming from this diagnostic is the basis for tomographic reconstructions that provide a 2D image of the plasma radiation profile. The reconstruction process itself is time-consuming. However, with a CNN trained on a large collection of sample tomograms, it becomes possible to produce those results much faster and with high accuracy.

In addition, the bolometer signals can be used to study disruption precursors. With a RNN trained on these signals, it is shown that bolometer data can provide a useful input for disruption prediction, both in terms of probability of disruption and time remaining to an impending disruption.

The paper is structured as follows. Section~\ref{sec:bolometry} provides a brief overview of the bolometer system, where the data is coming from, and of the tomographic method used at JET to reconstruct the plasma radiation profile. Section~\ref{sec:tomography} describes the CNN that has been developed for plasma tomography, and the results that can be obtained with it. Section~\ref{sec:disruptions} describes the RNN that has been developed for disruption prediction, and discusses its effectiveness when compared to other methods. Finally, Section~\ref{sec:conclusion} concludes the paper.

\section{The Bolometer System at JET}
\label{sec:bolometry}

One way to measure plasma radiation is through the use of bolometers, in particular foil bolometers~\cite{ingesson08tomography}. These sensors consist of a thin metal foil (about 10~$\mu$m) coupled with a temperature-sensitive resistor. As the metal foil absorbs radiation power, its temperature changes and there is a proportional change in resistance. Overall, such system provides a linear response to the absorbed power, in the range from ultraviolet (UV) to soft X-ray~\cite{mast91bolometer}.



The use of bolometers at JET dates back to the mid-1980s when the first bolometer system was installed~\cite{mast85bolometric}. This was followed by the development of other bolometer systems in the mid-1990s. One of these systems (KB3) was mainly directed at the divertor region at the bottom of the vessel. It has been subject to several improvements~\cite{ingesson00characterization}, and it is still partly in use today. Originally, KB3 included 7 cameras in the divertor region, each with 4 lines of sight.

\begin{figure}[h]
	\centering
	\includegraphics[scale=0.1]{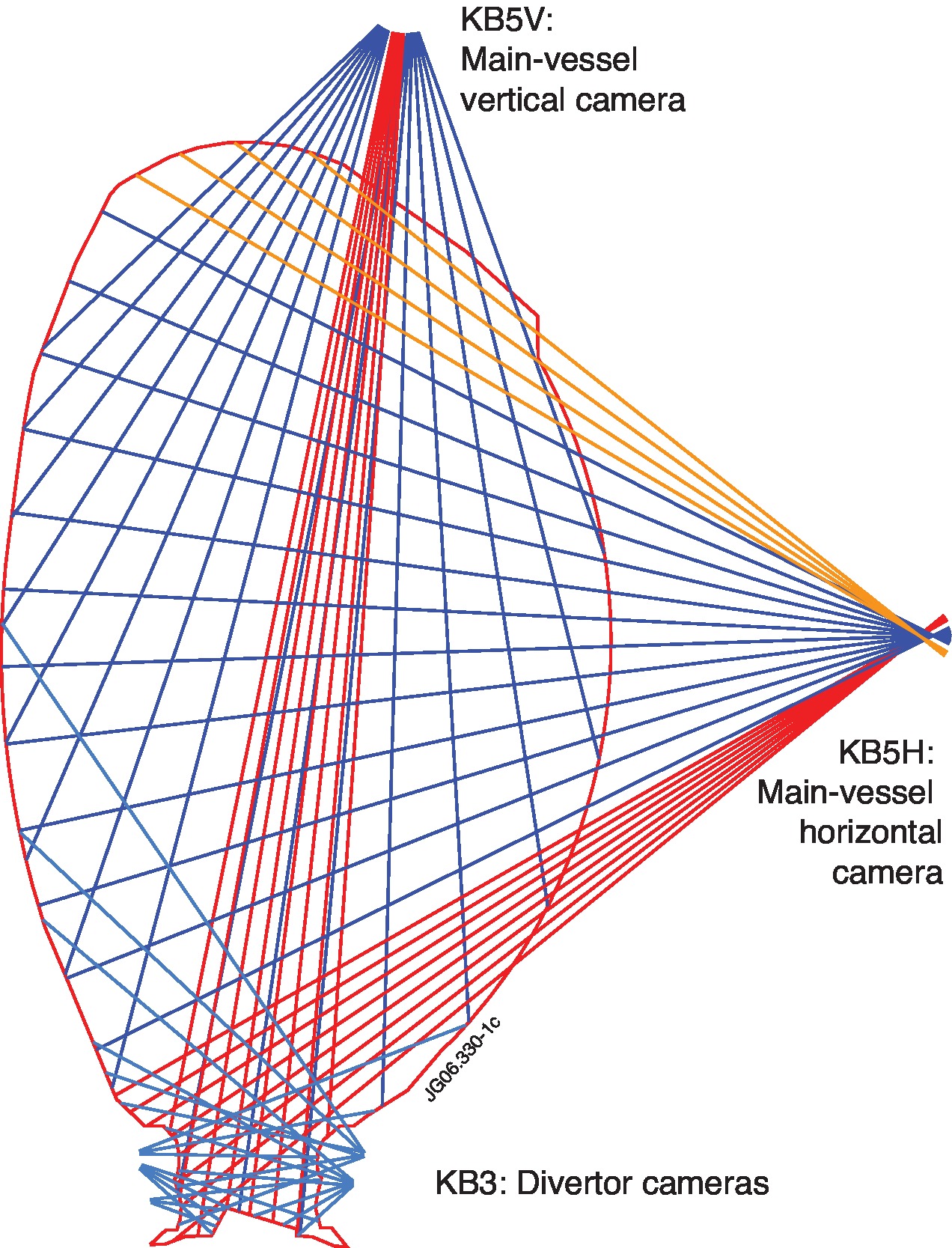}
	\caption{KB5 cameras and lines of sight. KB5H and KB5V are not in the same octant, so we assume that the plasma is toroidally symmetric. The lines of sight for KB3 are also shown. (EUROfusion figures database JG06.330-1c)}
	\label{fig:kb5}
\end{figure}

In the late 1990s and early 2000s, the current bolometer system (KB5) was developed \cite{mccormick05new,huber07upgraded,huber07improved}. This diagnostic comprises one horizontal camera and one vertical camera, with 24 bolometers each. The lines of sight for each of these cameras are arranged in such a way that 16 channels cover the whole plasma, and 8 channels are directed towards the divertor region, as illustrated in Figure~\ref{fig:kb5}. This means the system provides an overview of the plasma region, and also a more detailed view of the divertor region.

In addition, the vertical camera has 8 extra bolometers that can be used as reserve channels, so in total the system provides 24 + 24 + 8 = 56 lines of sight over the plasma.

Technically, each bolometer measures the line-integrated radiation along its line of sight. From the bolometer measurements, it is possible to reconstruct the 2D plasma radiation profile using tomography techniques. The underlying principle is the same as computed tomography (CT) in medical applications~\cite{buzug08book}, but the reconstruction method is different due to the scarce number of lines of sight available.

\begin{figure}[h]
	\centering
	\includegraphics[scale=0.4]{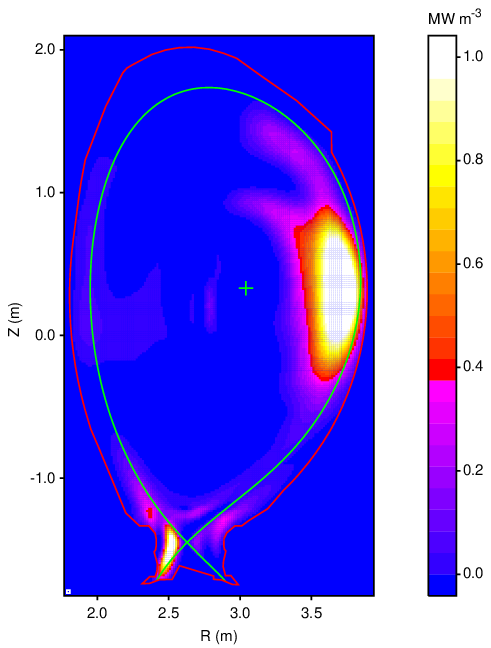}
	\caption{Example of a tomographic reconstruction for pulse 92213 at $t$=50.0s. Many of such reconstructions have been used as training data in this work.}
	\label{fig:tomo}
\end{figure}

The method that is used at JET~\cite{ingesson98tomography} actually predates the current bolometer system, and has been used with previous generations of soft X-ray diagnostics. In essence, it uses an iterative constrained optimization algorithm that minimizes the error with respect to the observed measurements, while requiring the solution to be non-negative. To do so, it solves a generalized eigenvalue problem in order to find a solution as a function of Lagrange multipliers, and then adjusts these Lagrange multipliers iteratively until the non-negativity constraints are satisfied~\cite{fehmers98algorithm}. Figure~\ref{fig:tomo} shows a sample result.

\begin{figure*}[b]
	\centering
	\includegraphics[scale=0.55]{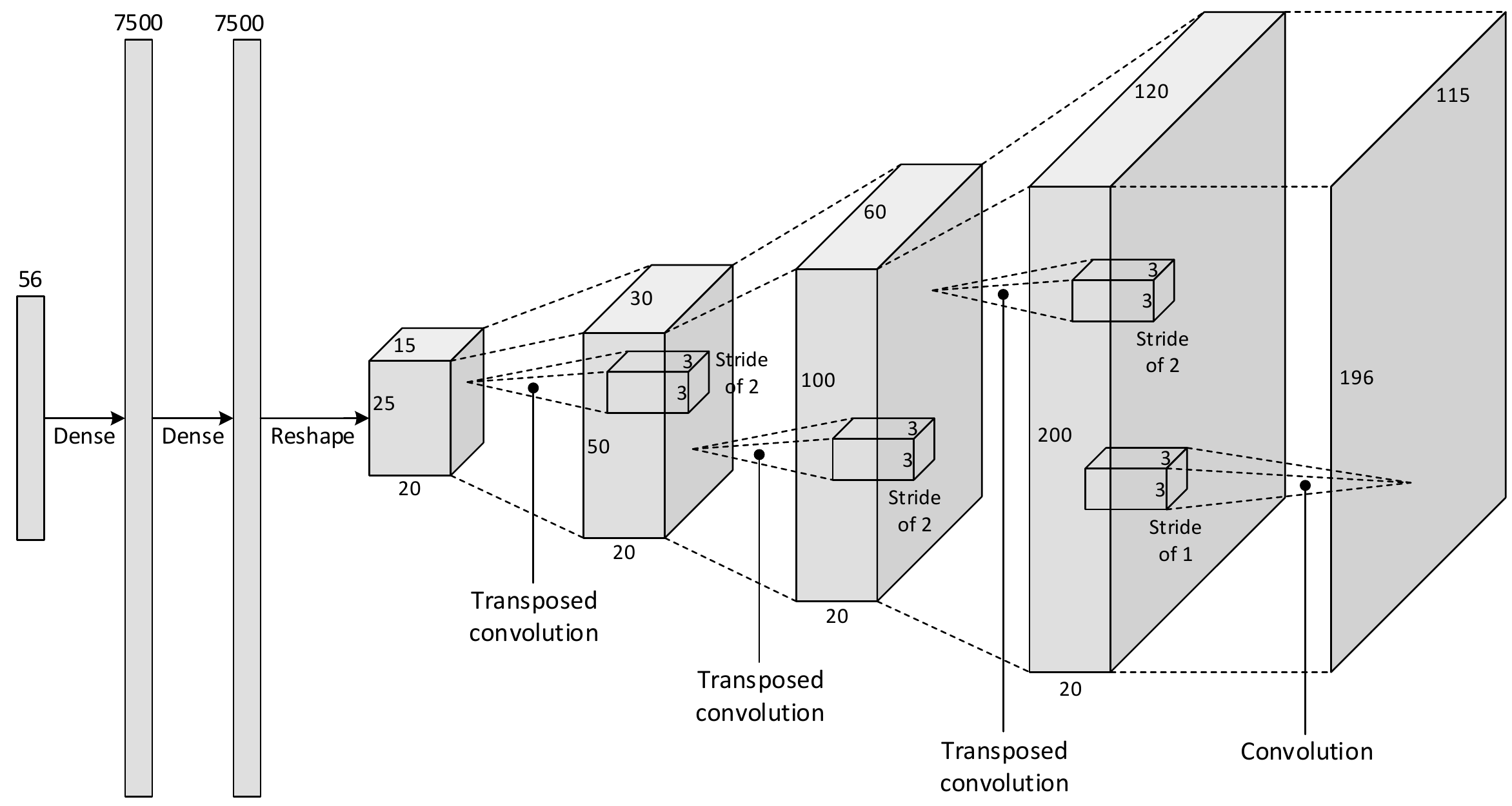}
	\caption{Deconvolutional neural network for plasma tomography. The input layer receives all the 56 channels from the bolometer system and the output layer provides an image with 196$\times$115 resolution.}
	\label{fig:cnn}
\end{figure*}

This iterative method takes a significant amount of computation time. The total run-time depends on the actual data, but it can take from several minutes up to 1 hour to produce a single reconstruction. Given that the bolometer system has a sampling rate of 5~kHz, and a JET pulse typically lasts for about 30 seconds, it could take several years to compute all the tomographic reconstructions for a single pulse.

In the next section, we describe how deep learning was used to train a CNN that is able to compute 3000 reconstructions per second on a Graphics Processing Unit (GPU).

\section{Deep Learning for Plasma Tomography}
\label{sec:tomography}

In general, CNNs have a structure comprising multiple convolutional layers. Each convolutional layer applies several filters (in the form of a small kernel or sliding window) to the input. The purpose of a filter is to detect a specific feature, so its output is called a \emph{feature map}. When CNNs are used for image classification, the first convolutional layer operates directly on the input image, while subsequent layers operate on the feature maps produced by previous layers.

In addition to the convolutional layers, CNNs contain subsampling layers. There is usually one subsampling layer after each convolutional layer. The purpose of having subsampling is to make the feature maps smaller and allow their number to progressively increase after each convolution, while keeping the network under a manageable size.

The output from these convolutional and subsampling layers is usually a large number of small feature maps. These feature maps are then flattened and connected to a couple of dense layers. The main idea behind image classification with CNNs is to have a first stage of convolutional layers to extract meaningful features from the input image, and a second stage of dense layers to perform the actual classification based on those features~\cite{lecun98gradient}.

In summary, the input to a CNN is typically a 2D image and its output is a 1D vector of class probabilities. However, for the purpose of plasma tomography, it would be useful to have a network that takes a 1D vector of bolometer measurements as input, and produces a 2D image of the plasma radiation profile as output. In the literature, the inverse of a CNN has been referred to as a \emph{deconvolutional network}, and it has found applications in image segmentation~\cite{noh15learning} and object generation~\cite{dosovitskiy15learning}. For example, by specifying the class label and the camera position (1D data), a deconvolutional network is able to generate an object (2D image) of the specified class from the given camera view~\cite{dosovitskiy17learning}.

\subsection{Network Architecture}

Figure~\ref{fig:cnn} shows the deconvolutional network developed for plasma tomography. It receives the bolometer measurements (a total of 56 lines of sight from both cameras) and produces a reconstruction of the plasma radiation profile with the same resolution as the tomographic reconstructions that are routinely produced at JET from bolometer data (196$\times$115 pixels).

After the network input, there are two dense layers with 7500 nodes, which are reshaped into a 3D tensor of size 25$\times$15$\times$20. This shape can be interpreted as comprising 20 features maps of size 25$\times$15. By applying a series of transposed convolutions, the feature maps are brought up to a size of 200$\times$120, from which the output image is generated by one last convolution. (The output shape of 200$\times$120 is trimmed to 196$\times$115 by a simple slicing operation.)

Essentially, a \emph{transposed convolution} is the inverse of a regular convolution in the sense that, if a sliding window would be applied to the output, the result would be the feature map given as input. It can be shown that learning a transposed convolution is equivalent to learning a weight matrix that is the transpose of a regular convolution, hence the name of this operation~\cite{dumoulin16guide}.

In Figure~\ref{fig:cnn}, the upsampling of feature maps from 25$\times$15 up to 200$\times$120 is achieved by having each transposed convolution operate with a stride of two pixels (i.e.~one pixel is being skipped between each two consecutive positions of the sliding window). This means that the output is four times larger than the input, except for the very last convolution which uses a stride of one to keep the same size.

\subsection{Network Training}

To train the network shown in Figure~\ref{fig:cnn}, we gathered all the tomographic reconstructions that have been computed at JET for the experimental campaigns from 2011 (since the installation of the ITER-like wall~\cite{matthews11ilw}) to 2016 (the time of the last completed campaign). We found 1219 pulses with a total of 27894 reconstructions (an average of about 23 reconstructions per pulse), and we split these pulses into 80\% for training, 10\% for validation, and 10\% for testing. The splitting was done by pulse in order to avoid having (possibly similar) reconstructions from the same pulse across the training, validation and test sets. Such splitting yielded a training set of 976 pulses with 22236 reconstructions, a validation set of 122 pulses with 2736 reconstructions, and a test set of 121 pulses with 2922 reconstructions.

\addtocounter{figure}{+1}
\begin{figure*}[b]
	\centering
	\includegraphics[width=\textwidth]{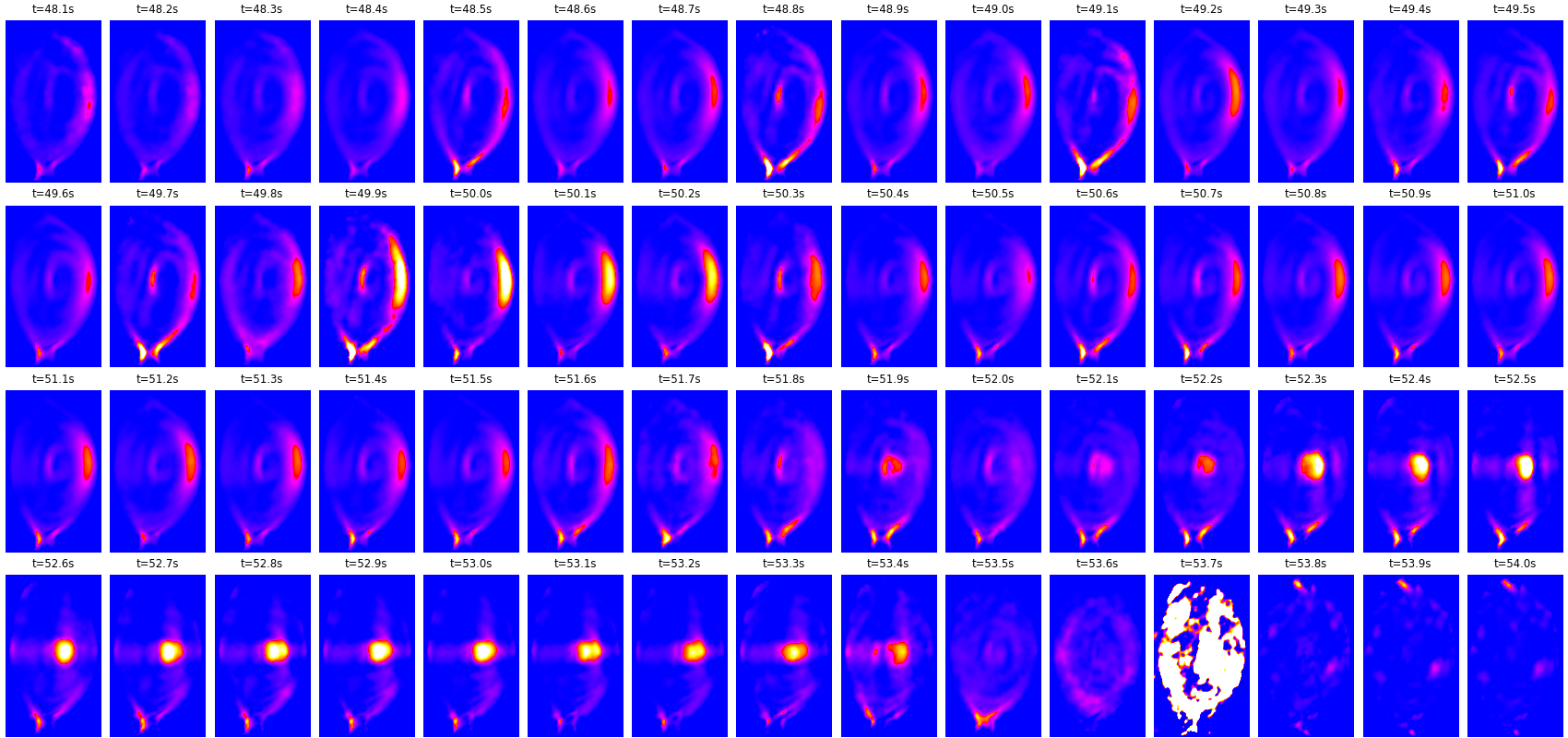}
	\caption{Reconstructions of pulse 92213 from $t$=48.1s to $t$=54.0s with a time step of 0.1s. A video of the reconstructions for this pulse can be found in the supplementary material online.}
	\label{fig:pulse}
\end{figure*}

The network was trained on an NVIDIA Titan X GPU using accelerated gradient descent (Adam~\cite{kingma14adam}) with a small learning rate ($\text{10}^{-\text{4}}$) and a batch size of 436 samples. The batch size was chosen to be a perfect divisor of the number of training samples (22236/436 = 51) in order to avoid having any partially filled batch in the training set.

The network was trained to minimize the mean absolute error between the output and the sample tomograms that were provided for training. Figure~\ref{fig:loss_cnn} shows the evolution of the training loss and of the validation loss across 1600 epochs. The best result was achieved at around epoch 800, with a minimum validation loss of 0.0128~MW~m$^{-\text{3}}$. To appreciate the small scale of this error, it can be compared to the dynamic range of the reconstruction in Figure~\ref{fig:tomo} (1.0 MW m$^{-\text{3}}$).

\addtocounter{figure}{-2}
\begin{figure}[h]
	\centering
	\includegraphics[scale=0.5]{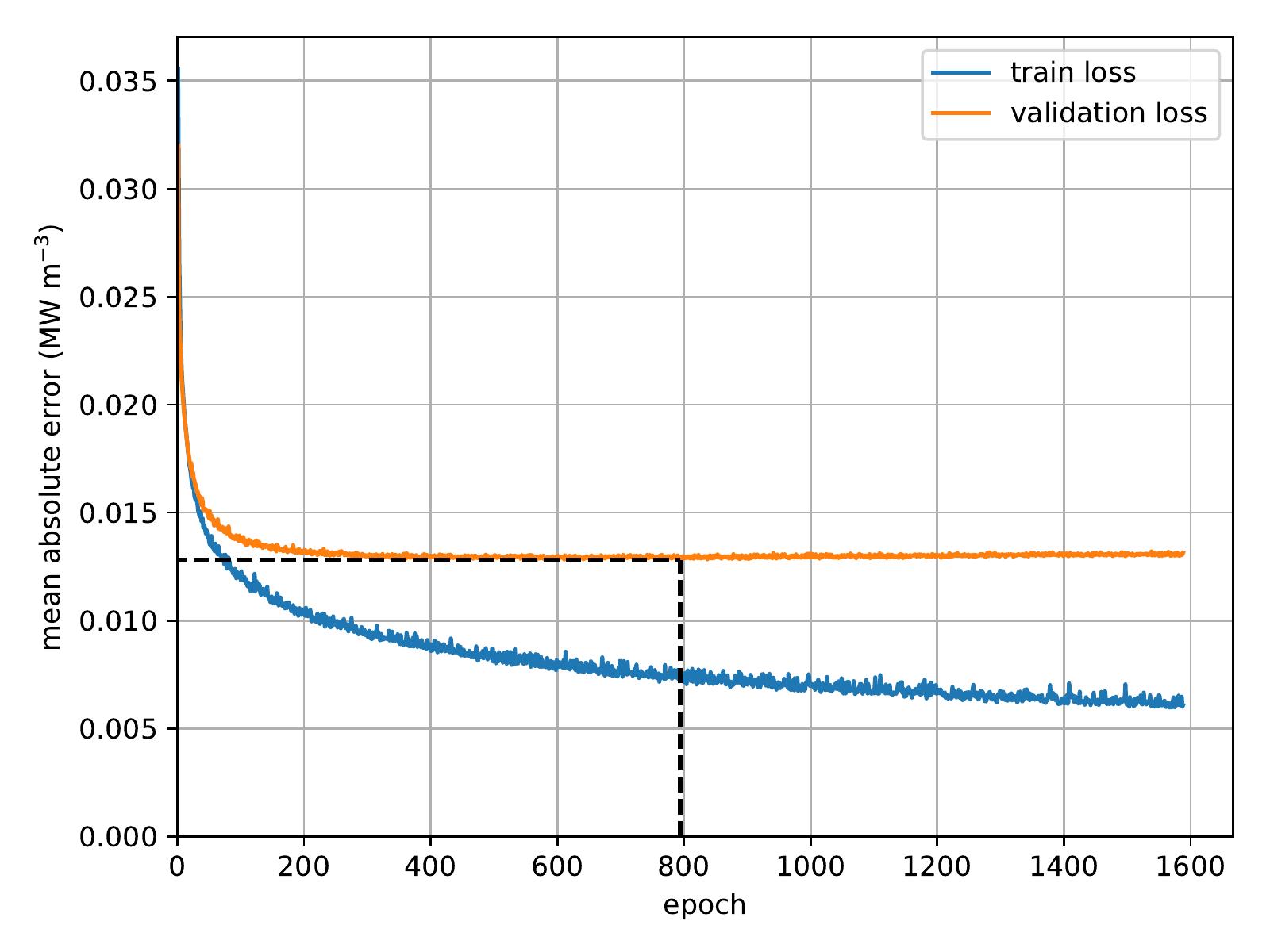}
	\caption{Loss and validation loss during training of the deconvolutional network. The minimum validation loss, and the epoch at which it occurred, are indicated by dashed lines.}
	\label{fig:loss_cnn}
\end{figure}
\addtocounter{figure}{+1}

To assess the quality of the reconstructions produced by the network, it is possible to use image comparison metrics, such as structural similarity (SSIM)~\cite{wang04ssim} or peak signal-to-noise ratio (PSNR)~\cite{thu08psnr}. On the test set, the network achieves an SSIM value of 0.936$\pm$0.061 (the maximum is 1.0) and a PSNR of 35.4$\pm$7.2 dB, which is roughly similar to the error introduced when compressing an image to JPEG format.

\subsection{Full-pulse reconstructions}

Once trained, the network can be used to generate the reconstruction for any given pulse at any given point in time. In fact, given the bolometer data from an entire pulse, the network can generate a batch of reconstructions for every point in time. This way, it is possible to analyze the evolution of the plasma radiation profile across a full experiment.

As an example, Figure~\ref{fig:pulse} shows the reconstruction of pulse 92213 from $t$=48.1s onwards, with a time step of 0.1s (due to space restrictions only). The first and second rows show a focus of radiation developing on the outer wall, which seems to slowly fade away (row 3), only to reappear later at the plasma core with particularly strong intensity (rows 3--4).

This radiation peaking stays at the core for a relatively long time (at least from $t$=52.3s to $t$=53.4s), while changing slightly in shape during that interval. Eventually, it also fades away as the heating systems are being turned off. However, just as it seemed that the plasma was about to soft land, there is a disruption at around $t$=53.7s.

\begin{figure}[h]
	\centering
	\includegraphics[scale=0.5]{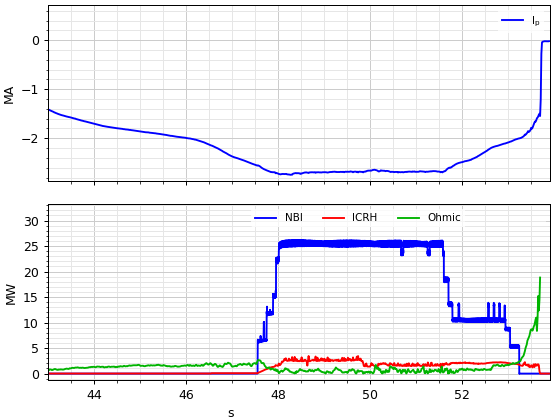}
	\caption{Plasma current (top) and heating systems (bottom) for pulse 92213, including NBI (neutral beam injection), ICRH (ion cyclotron resonance heating) and ohmic heating. }
	\label{fig:dashboard}
\end{figure}

For comparison, Figure~\ref{fig:dashboard} shows the plasma current and heating power for this pulse. The disruption is clearly marked by a sudden current quench.

Disruptions are one of the major problems affecting tokamaks today. While the dynamics of disruptions are not yet fully understood, one of the main reasons is believed to be impurity accumulation at the plasma core, which decreases the core temperature and eventually leads to core collapse~\cite{vries11survey}.

To some extent, this phenomenon is clearly visible in Figure~\ref{fig:pulse}. It is therefore not surprising that the bolometer system has played an key role in several disruption studies at JET~\cite{arnoux09heat,riccardo10disruption,huber11radiation,lehnen11disruption}, either by providing a measurement of total radiation, or by providing the tomographic reconstruction at a few specific points in time. With the proposed network, it becomes possible to quickly generate a series of reconstructions that provide a detailed view of how the plasma profile evolves across time.

\section{Deep Learning for Disruption Prediction}
\label{sec:disruptions}

Over the years, a number of different approaches to disruption prediction have been developed, including the use of neural networks~\cite{pautasso02online,cannas04disruption,windsor05cross,yoshino05neural}, support vector machines~\cite{cannas07svm,ratta10advanced,lopez14implementation} and decision trees~\cite{murari08prototype,murari09unbiased}, among other techniques. In all of these approaches, the prediction model takes a set of global plasma parameters as input (e.g.~the plasma current, the locked mode amplitude, the total input power, the safety factor, the poloidal beta, etc.) and outputs a probability of disruption, or the remaining time to a predicted disruption.

More recently, the use of random forests~\cite{rea18exploratory,rea18investigations} and even deep learning~\cite{harbeck19predicting,sharma18proposed} have shown promising results, but the input signals that are provided to these models still consist of about ten global plasma parameters derived from different diagnostics.

In the previous section, we have seen that a bolometer system with multiple lines of sight is able to capture some of the physical phenomena associated with plasma disruptions. From such bolometer data, it is possible to reconstruct the internal shape of the plasma radiation profile and observe, for example, an impurity concentration at the plasma core, which could be the precursor to an impending disruption.

\begin{figure}[h]
	\centering
	\includegraphics[scale=0.5]{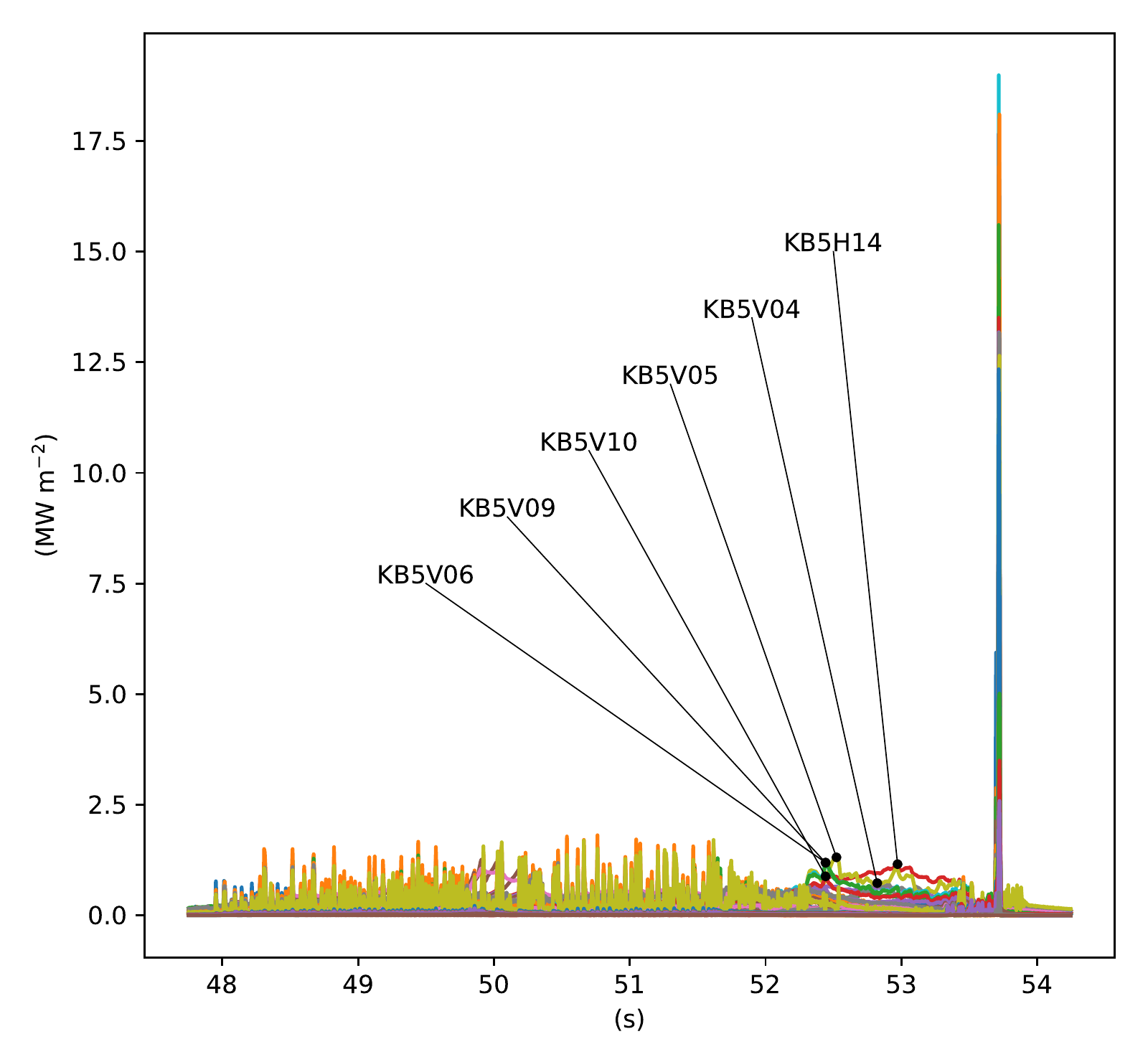}
	\caption{Bolometer signals for pulse 92213. The channels that become more active before the disruption are indicated in the figure.}
	\label{fig:bolo}
\end{figure}

This suggests that the sensor measurements coming from the bolometer system could, in principle, be used for disruption prediction. Figure~\ref{fig:bolo} shows the bolometer signals for the same pulse that was reconstructed in Figure~\ref{fig:pulse}. The disruption is clearly visible at around $t$=53.7s and it would be easy to detect it from these signals. However, the point is whether it is be possible to predict the disruption before it occurs.

As can be observed in Figure~\ref{fig:bolo}, before the disruption there is a period of about one second where the bolometer signals display a distinct pattern from the rest of the pulse. This corresponds to the radiation peaking at the plasma core. The bolometers that are yielding the strongest signals (e.g. KB5V channel 9, KB5V channel 5, and KB5H channel 14) are pointing precisely to that region.

In general, any model that is used for disruption prediction should be able to detect these or other kinds of precursors from data. In previous works, most authors used a set of global plasma parameters to learn and detect disruption precursors. Here, we will be using the bolometer signals, which provide a spatial view of the plasma radiation profile.

In addition to the spatial view, another dimension that will be important to consider is time. As shown in Figure~\ref{fig:bolo}, the disruption precursor develops and endures for a relatively long time (on the order of a second). Therefore, it will be important to consider not only the bolometer measurements at a certain point in time, but also the preceding measurements in order to have a sense of how the plasma behavior is developing and how far it is from a possible disruption.

With both dimensions, the input to the predictor will be a sequence of vectors, where each vector contains the bolometer measurements at a given point in time. This closely resembles the way in which RNNs are used in natural language processing. In that case, the input is also a sequence of vectors, where the sequence is a text sentence and each vector is a word mapped into a vector space of a certain dimensionality, through a process called \emph{word embedding}~\cite{goldberg17neural}.

\subsection{Network Architecture}

\begin{figure*}[b]
	\centering
	\includegraphics[scale=0.65]{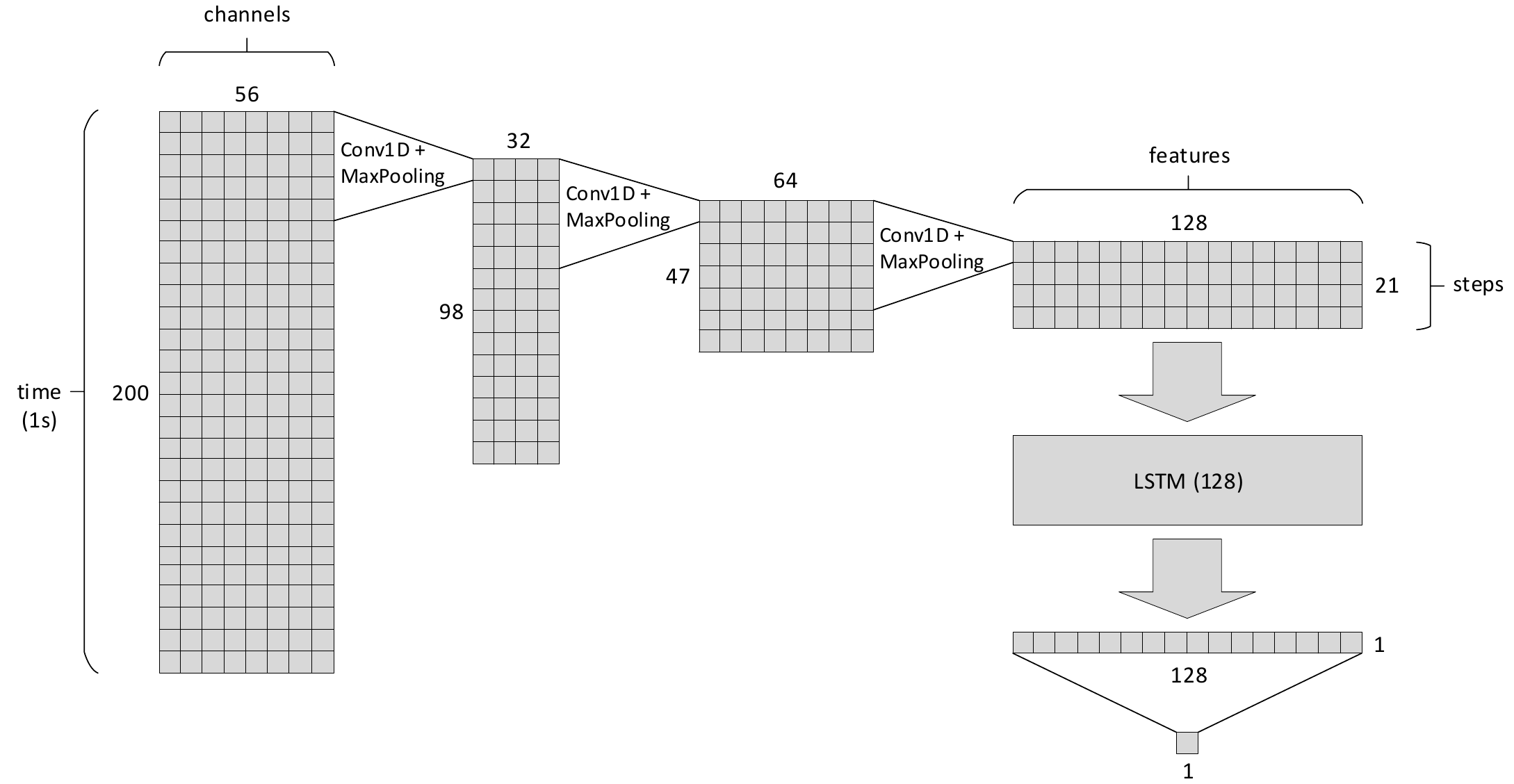}
	\caption{Recurrent neural network for disruption prediction. The input layer receives a full second of data (200 samples) from the bolometer signals (56 channels) and the output layer provides a single scalar prediction, which can be either the time-to-disruption or the probability of disruption.}
	\label{fig:rnn}
\end{figure*}

A RNN is able to handle sequences of vectors by processing each vector at a time. The key distinctive feature of RNNs is that they maintain an internal state, and this internal state is fed back to the input along with the next vector in the sequence. A RNN can therefore ``remember'' features from past vectors in the sequence.

In simple RNNs, such ``memory'' is necessarily short-lived, since the network keeps updating its internal state as it receives new vectors, and gradually ``forgets'' about vectors that are in the distant past. To provide RNNs with the ability to maintain long-term dependencies, more advanced architectures have been proposed, namely the \emph{long short-term memory} (LSTM)~\cite{hochreiter97lstm,gers02learning}. In a LSTM, information can be carried across multiple time steps, and that information can be updated or forcefully forgotten at each time step, through the use of an \emph{input gate} and a \emph{forget gate}, respectively.

More recently, the combination of CNNs and RNNs has been shown to be very effective in tasks such as video processing~\cite{donahue15longterm,donahue17longterm} and text classification~\cite{wang16dimensional,wang16coling}. In essence, the CNN works as a feature extractor before passing the data to a RNN for sequential processing.

In this work, we use a similar approach to preprocess the bolometer data through a series of 1D convolutional layers before handing them over to a LSTM. As in a traditional CNN, we use subsampling after each convolutional layer to progressively reduce the length of the time sequence, while increasing the number of filters that are applied to it.

The resulting network is shown in Figure~\ref{fig:rnn}. The input has a shape of 200$\times$56, with the first being the sequence length, and the latter being the vector size. The sequence length of 200 time steps correspond to a 200 Hz downsampling of the bolometer signals. With a sequence of 200 time steps at 200 Hz, this means that one second of bolometer data is being considered in order to produce a prediction.

As before, the vector size of 56 corresponds to the number of lines of sight available in the bolometer system. Some of these lines correspond to unused reserve channels, which yield always a measurement of zero. Others correspond to bolometers which are known to be faulty, providing erratic measurements. In these cases, the network is expected to filter out those channels, so the number of filters used in the first convolutional layer (32) is actually less than the number of input channels (56), as shown in Figure~\ref{fig:rnn}.

In subsequent layers, the number of filters doubles after each convolution. A kernel size of 5 time steps is used in all convolutions, together with a max-pooling operation in order to reduce the sequence length to half. Eventually, the convolutional part of the network yields a sequence of 21 steps, where each step contains a 128-feature vector.

This sequence of feature vectors is handed over to a LSTM with 128 units, which returns the last output after having processed the whole sequence. One final dense layer combines these vector elements into a single output. This output can be either the probability of disruption or the time to disruption, as described next.

\subsection{Network Training}

The network can be trained for two different objectives:
\begin{itemize}
	\setlength\itemsep{0.5em}
	
	\item \emph{Classification} -- To predict the probability of disruption (i.e.~the probability that the current pulse will end in a disruption), the last layer of the network will contain a sigmoid activation to produce a value between 0 and 1. The loss function used to train the network will be the binary cross-entropy~\cite{goodfellow16book}, and the network will be trained using both disruptive and non-disruptive pulses.
	
	\item \emph{Regression} -- To predict the time to disruption (i.e.~the remaining time up to an incoming disruption in the current pulse), the last layer of the network will have a linear activation, so that it can produce positive as well as negative values within any range. The loss function used to train the network is the mean absolute error, and the network is trained using disruptive pulses only.
	
\end{itemize}

To train these two variants, we gathered the bolometer signals for all pulses in the experimental campaigns from 2011 to 2016. In addition, we identified the disruptive pulses and gathered their disruption times from the JET disruption database~\cite{gerasimov18overview}. Pulses with deliberate (i.e. intentional) disruptions were excluded from the dataset, as is done in other works~\cite{vega13results}. Eventually, this yielded a total of 9323 pulses, 1444 of which were disruptive.

Again, we divided these pulses into 80\% for training, 10\% for validation, and 10\% for testing. To train the network for probability of disruption, we used a data generator to draw samples at random time points from each pulse and feed those samples to the network, together with a binary label to indicate whether the sample was drawn from a disruptive or non-disruptive pulse. To train the network for time to disruption, we used a data generator to draw samples from disruptive pulses only, together with a label to indicate the time interval between the sample time and the disruption time.

In both cases, a batch size of 500 samples was used, where each sample is a 200$\times$56 sequence of vectors. The sample time is defined as the latest time step in that sequence. Each training epoch was defined as a run through 50 batches. For validation purposes, we used samples drawn at regular intervals from the validation pulses.

\begin{figure}[t]
	\centering
	\includegraphics[scale=0.5]{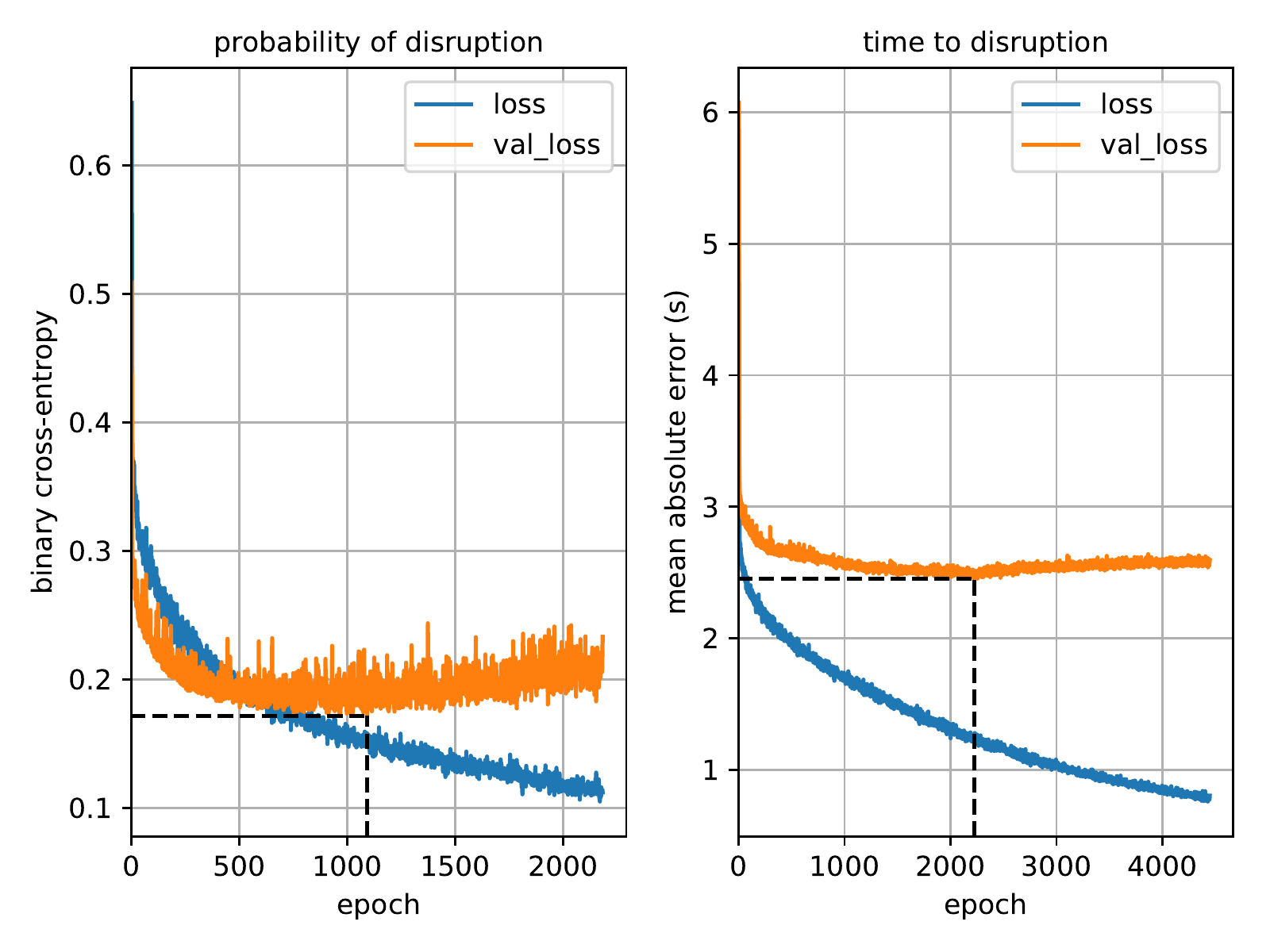}
	\caption{Loss and validation loss during training of the recurrent network. The minimum validation loss, and the epoch at which it occurred, are indicated by dashed lines.}
	\label{fig:loss_rnn}
\end{figure}

Figure~\ref{fig:loss_rnn} shows the training loss and the validation loss for both variants. For probability of disruption, training ran for 2200 epochs, with the best result being achieved at around epoch 1100 with a validation loss (binary cross-entropy) of 0.172. For time to disruption, training ran for 4400 epochs with the best result being achieved at around epoch 2200 with a validation loss (mean absolute error) of 2.452 seconds.

Although these loss values may appear to be somewhat high, it should be noted that they are averaged over many samples. For samples that are far away from the disruption, the loss will be high because it is difficult to predict when or if a disruption will occur. This will be compensated by a small loss for samples that are near the disruption. Thus, the prediction will be more accurate when it matters the most.

In particular, the validation loss for the probability of disruption (left chart in Figure~\ref{fig:loss_rnn}) seems to be quite noisy. This has to do with the fact that random samples are being drawn from both disruptive and non-disruptive pulses. In some cases (e.g. at the beginning of a pulse) samples from disruptive and non-disruptive pulses will be similar, despite having opposite labels. This makes the network weights oscillate across training in order to meet such contradictory labels.

\subsection{Full-pulse predictions}

\begin{figure}[t]
	\centering
	\includegraphics[scale=0.5]{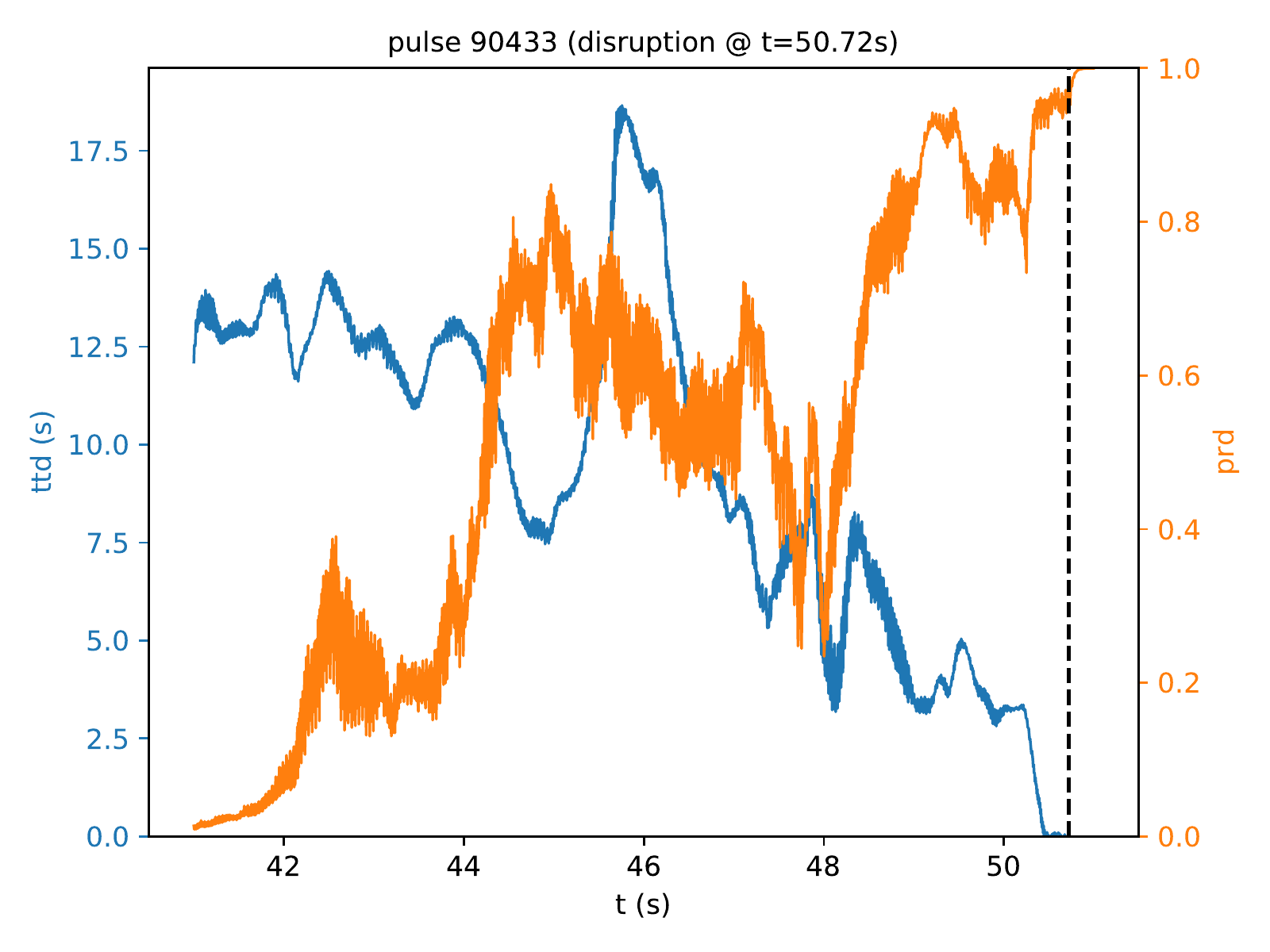}
	\caption{Time to disruption (\emph{ttd}) and probability of disruption (\emph{prd}) for a disruptive pulse. The time of disruption is indicated by a vertical dashed line.}
	\label{fig:pred}
\end{figure}

\begin{figure}[t]
	\centering
	\includegraphics[scale=0.5]{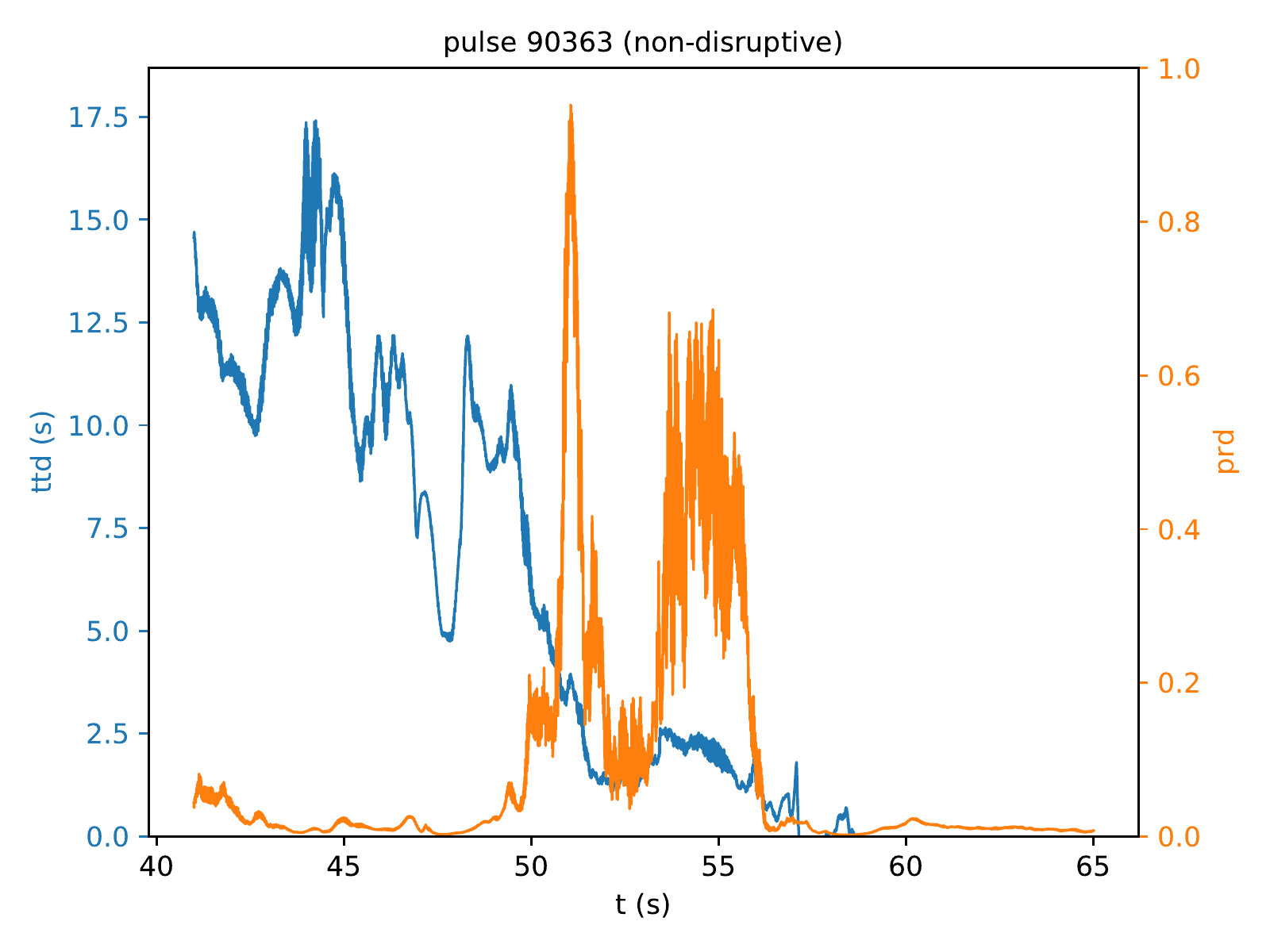}
	\caption{Time to disruption (\emph{ttd}) and probability of disruption (\emph{prd}) for a non-disruptive pulse. The time-to-disruption reaches zero, even though there is no disruption.}
	\label{fig:pred_2}
\end{figure}

To illustrate the results that can be obtained with the trained network, Figure~\ref{fig:pred} shows the step-by-step predictions of the probability of disruption and of the time to disruption on an entire pulse. The disruption time is marked with a vertical dashed line at $t$=50.72s. For comparison, Figure~\ref{fig:pred_2} shows the same predictions for a non-disruptive pulse.

Some important observations can be made from these and other similar results:
\begin{itemize}

	\item The disruption is characterized by a point in time where the probability of disruption is close to 1.0 and the time to disruption is close to 0.0. Therefore, when the probability of disruption approaches 1.0 and the time to disruption approaches 0.0 at the same time, this is a strong indicator that a disruption is possibly imminent.

	\item It often happens that the probability of disruption is high at some intermediate points during the pulse, even for non-disruptive pulses. However, such indicator alone is not meaningful if the time to disruption is still high.
	
	\item It also happens that the time to disruption reaches zero for non-disruptive pulses. (Since this variant of the network has been trained exclusively on disruptive pulses, it will keep predicting that a disruption will occur even for non-disruptive pulses.) However, such indicator alone is not meaningful if the probability of disruption is low.
	
\end{itemize}

From these observations, it follows that the probability of disruption and the time to disruption should both be taken into account when deciding to take some action. For example, when the probability of disruption exceeds a certain threshold and the time to disruption falls below another threshold, an alarm could be raised to trigger the \emph{disruption mitigation valve} (DMV)~\cite{finken11dmv} which, in a matter of a few milliseconds, will inject a massive amount of gas to cool down the plasma.\footnote{An alternative approach that is currently being developed for disruption mitigation at JET and other tokamaks is \emph{shattered pellet injecton}~\cite{baylor18shattered}.}

However, even when using both predictors, there might be some false alarms. For example in Figure~\ref{fig:pred_2}, at around $t$=51s, the probability of disruption rises and the time to disruption decreases in a way that an alarm could be triggered, even though the pulse is non-disruptive. Also, there might be situations where the network might not be able to predict a disruption on time (missed alarms).

\subsection{Prediction accuracy}

As with any binary classification problem, there will be true positives (successfully predicted disruptions), true negatives (correctly predicted absence of disruption), false positives (false alarms), and false negatives (missed alarms). Table~\ref{tab:thresholds} shows how the alarm-triggering thresholds on probability of disruption and on time to disruption affect the prediction results on the test set, which includes 932 pulses. For convenience, the results are being reported in terms of percentage. As a reference, the test set contains 157 (16.85\%) disruptive pulses, and 775 (83.15\%) non-disruptive pulses.

\begin{table}[h]
	\def\arraystretch{1.1}
	\setlength\tabcolsep{1.5mm}
	\scriptsize
	\centering
	\caption{Prediction results on the test set with different alarm-triggering thresholds}
	\label{tab:thresholds}
	\begin{tabular}{c|c|c|c|c}
		& & & & \\
		\& & \emph{prd} $\geq$ 0.95 & \emph{prd} $\geq$ 0.90 & \emph{prd} $\geq$ 0.85 & \emph{prd} $\geq$ 0.80\\
		& & & & \\
		\hline
		\emph{ttd} $\leq$ 0.5 s & \textbf{\cell{7.73}{80.58}{2.58}{9.12}} & \cell{9.66}{79.83}{3.33}{7.19} & \cell{10.62}{79.51}{3.65}{6.22} & \cell{11.27}{78.54}{4.61}{5.58} \\
		\hline
		\emph{ttd} $\leq$ 1.0 s & \cell{8.15}{80.26}{2.90}{8.69} & \textbf{\cell{9.98}{79.51}{3.65}{6.87}} & \cell{11.16}{78.97}{4.18}{5.69} & \cell{12.02}{76.93}{6.22}{4.83} \\
		\hline
		\emph{ttd} $\leq$ 1.5 s & \cell{9.12}{79.94}{3.22}{7.73} & \cell{10.84}{78.97}{4.18}{6.01} & \textbf{\cell{11.70}{77.90}{5.26}{5.15}} & \cell{12.34}{76.29}{6.87}{4.51} \\
		\hline
		\emph{ttd} $\leq$ 2.0 s & \cell{9.55}{79.40}{3.76}{7.30} & \cell{11.59}{78.22}{4.94}{5.26} & \cell{12.34}{77.15}{6.01}{4.51} & \textbf{\cell{12.88}{75.54}{7.62}{3.97}} \\
	\end{tabular}
\end{table}

The results in bold (diagonal) illustrate the trade-off between false alarms (FP) and missed alarms (FN). As the threshold on probability of disruption becomes lower, and as the threshold on time to disruption becomes higher, the network is able to catch more disruptions, but at the expense of raising more false alarms. A good balance seems to be \emph{prd} $\geq$ 0.85 and \emph{ttd} $\leq$ 1.5, where both precision TP/(TP+FP) = 69.0\% and recall TP/(TP+FN) = 69.4\% yield a similar value.

These results are below those reported in the literature. For example, the disruption predictor that is currently being used at JET (APODIS) has a recall of 85.38\% and a false alarm rate (FP) of 2.46\%~\cite{moreno16disruption}. However, it should be noted that APODIS is based on a set of global plasma parameters, while our prediction is based on the bolometer signals alone. The conclusion is that the bolometer signals are informative for disruption prediction, and they could possibly be combined with other inputs in order to improve the accuracy of current predictors.

In recent works~\cite{pau18analysis,pau19approach} the bolometer signals have been used to extract one or two peaking factors of the radiation profile, which were then combined with other global parameters for disruption prediction. An advantage of using deep learning is that, in principle, it should be possible to extract these and/or other relevant features directly from the raw signals.

\subsection{Warning time}

Every time a disruption is correctly predicted, it is possible to calculate the \emph{warning time} as the difference between the disruption time and the time at which the alarm was triggered. For the alarm-triggering thresholds \emph{prd} $\geq$ 0.85 and \emph{ttd} $\leq$ 1.5 we have already seen that the recall (i.e. the percentage of correctly predicted disruptions) is TP/(TP+FN) = 69.4\%. The question now is how far ahead, in terms of warning time, those disruptions are predicted.

\begin{figure}[h]
	\centering
	\includegraphics[scale=0.5]{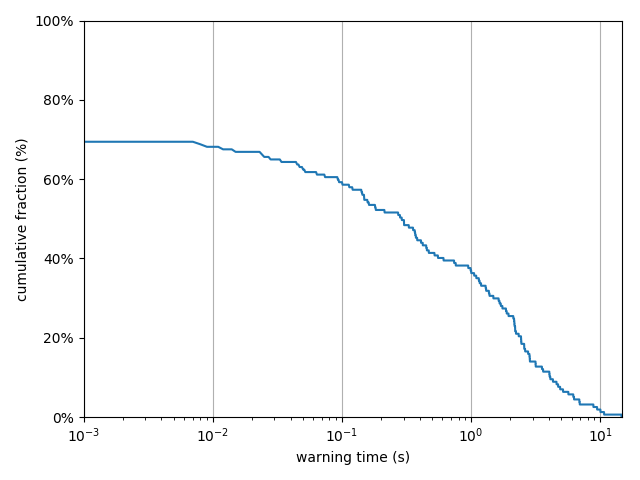}
	\caption{Cumulative fraction of correctly predicted disruptions vs.~warning time, for a recurrent neural network based on bolometer data.}
	\label{fig:warning_time}
\end{figure}

Figure~\ref{fig:warning_time} shows the cumulative fraction of correctly predicted disruptions as a function of warning time. As expected, it is more difficult to predict disruptions far away from the disruption time (i.e. with higher warning times). However, in global terms, this yields an average warning time of about 2~s, which is significantly higher than that of APODIS (0.35~s). This suggests that a recurrent network applied to the time series of bolometer data may contribute to detecting disruption-relevant behavior earlier than other methods.

\section{Conclusion}
\label{sec:conclusion}

In this work, we have applied deep learning to a single, multi-channel diagnostic at JET. Based on the signals from this diagnostic, we used a CNN to reconstruct the plasma radiation profile, and a RNN to perform disruption prediction. The CNN provides results with high accuracy, and is therefore a reliable alternative to more time-consuming methods. On the other hand, the RNN does not reach the performance of other methods presently used at JET, but it provides the prospect of improving current disruption predictors.

Both of these tasks could benefit from the use of data from multiple diagnostics. For example, it is possible to use the magnetic equilibrium to improve the plasma radiation profile, and this could be provided as an additional input to the CNN. On the other hand, disruption prediction can be made more accurate by taking into account several plasma parameters, and these could be additional inputs to the RNN.

In the future, it should be possible to build large deep learning models using data from many diagnostics in order to gain insights into the plasma behavior. The approaches described herein are just the beginning, and it is possible to envision that these approaches can be extended and applied to other fusion devices with minimal adaptations.

\section*{Acknowledgments}

This work has been carried out within the framework of the EUROfusion Consortium and has received funding from the Euratom research and training programme 2014-2018 and 2019-2020 under grant agreement No 633053. The views and opinions expressed herein do not necessarily reflect those of the European Commission. IST activities also received financial support from \textit{Funda\c{c}\~{a}o para a Ci\^{e}ncia e Tecnologia} through project UID/FIS/50010/2019. The Titan X GPU used in this work was donated by NVIDIA
Corporation.

\bibliographystyle{IEEEtran}
\bibliography{paper}





\end{document}